\documentclass[preprint,showpacs,preprintnumbers,amsmath,amssymb]{revtex4}
\usepackage{amsmath}

\usepackage{graphicx}
\usepackage{pdfpages}
\usepackage{dcolumn}
\usepackage{bm}
\usepackage{amsmath} 
\usepackage{amsfonts}
\usepackage{amssymb}
\usepackage{url}
\usepackage{microtype}
\usepackage{graphicx}%

\newcommand{\qed}{\nobreak \ifvmode \relax \else
      \ifdim\lastskip<1.5em \hskip-\lastskip
      \hskip1.5em plus0em minus0.5em \fi \nobreak
      \vrule height0.75em width0.5em depth0.25em\fi}

\begin{document}

\preprint{}
\title{Ordinal and Cardinal Dendrograms Depicting Migration-Based Regionalization of 3,000 + U. S. Counties}
\author{Paul B. Slater}
 \email{slater@kitp.ucsb.edu}
\affiliation{%
Kavli Institute for Theoretical Physics, University of California, Santa Barbara, CA 93106-4030\\
}
\date{\today}
\begin{abstract}
We have obtained a ``hierarchical regionalization" of 3,107 county-level units of the United States based upon census-recorded 1995-2000 intercounty migration flows. The methodology employed was the two-stage (double-standardization and strong component [directed graph] hierarchical clustering) algorithm described in the 2009 PNAS (106 [26], E66) letter (arXiv:0904.4863).
Various features  (e. g., cosmopolitan vs. provincial aspects, and indices of isolation) of the regionalization have been previously discussed in 
arXiv:0907.2393, arXiv:0903.3623 and arXiv:0809.2768. However, due to the lengthy (38-page) nature of the associated dendrogram, the detailed tree structure itself was not readily available for inspection. Here, we do present this (county-searchable) dendrogram--and invite readers to explore it, based on their particular interests/locations. An ordinal scale--rather  than the originally-derived cardinal
scale of the doubly-standardized values--in which groupings/features were more immediately apparent, was originally presented. Now, we append the  cardinal-scale dendrogram. 
\end{abstract}
\maketitleƒ

\section{Introduction}
The principal purpose of this report is to present--in conveniently accessible (county-searchable) form--two lengthy (3,107-unit, 38-page) dendrograms based upon 1995-2000 U. S. intercounty migration flows. These dendrograms had underpinned the discussion in our earlier studies 
\cite{SlaterDendrogram,Multiscale,MatrixPlots}, but had been omitted there for brevity's sake. (We attempted to include a dendrogram as an EPAPS file in 
\cite{Multiscale}, but it seems challenging to retrieve.) The methodology (two-stage--double-standardization and strong component [directed graph] hierarchical clustering \cite{carlsson2014hierarchical,tarjan1983improved}) employed to yield the dendrograms had been briefly discussed--including its earlier widespread applications to various forms of "transaction flows"--in the PNAS letter \cite{PBSPNAS}. Extensive bibliographies indicating these diverse applications over the years are available in \cite{SlaterDendrogram,Multiscale,MatrixPlots,PBSPNAS}.

Here, to begin, we present the 38-page long dendrogram on an ordinal scale (sec.~\ref{Ordinal}), followed by an abbreviated discussion of its multitudinous properties (sec.~\ref{Discussion}). Then, the original cardinal-scale dendrogram will be shown, as well (sec.~\ref{Cardinal}).

We have earlier still applied the same methodology to the (thirty-years previous) 1965-70 U. S. intercounty migration flows. Studies which concern various aspects of that earlier analysis are \cite{MigrationRegions,slater1984partial,slater1984origin,slater1984measuring,slater1987algorithm}.

\section{Ordinal-scale Dendrogram} \label{Ordinal}
\includegraphics[page=1,scale=0.95]{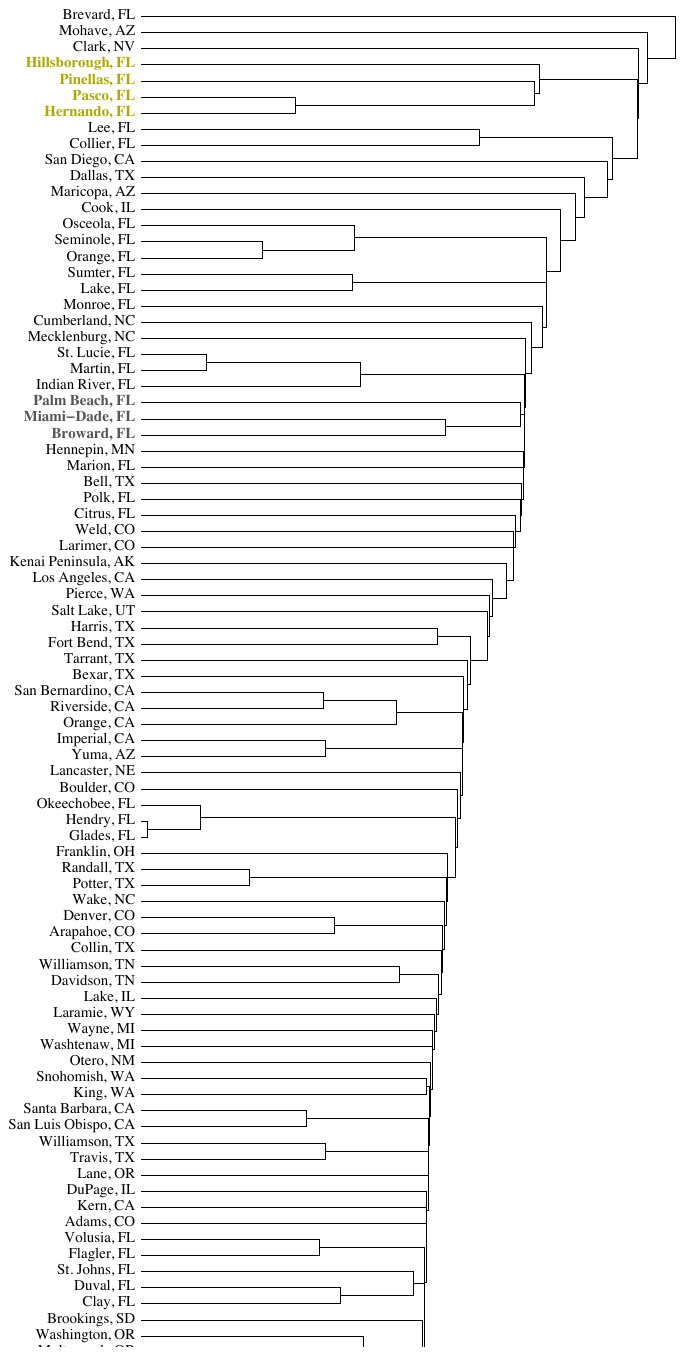}

\includegraphics[page=2,scale=0.95]{USCountyHierarchy1.pdf}

\includegraphics[page=3,scale=0.95]{USCountyHierarchy1.pdf}

\includegraphics[page=4,scale=0.95]{USCountyHierarchy1.pdf}

\includegraphics[page=5,scale=0.95]{USCountyHierarchy1.pdf}

\includegraphics[page=6,scale=0.95]{USCountyHierarchy1.pdf}

\includegraphics[page=7,scale=0.95]{USCountyHierarchy1.pdf}

\includegraphics[page=8,scale=0.95]{USCountyHierarchy1.pdf}

\includegraphics[page=9,scale=0.95]{USCountyHierarchy1.pdf}

\includegraphics[page=10,scale=0.95]{USCountyHierarchy1.pdf}

\includegraphics[page=11,scale=0.95]{USCountyHierarchy1.pdf}

\includegraphics[page=12,scale=0.95]{USCountyHierarchy1.pdf}

\includegraphics[page=13,scale=0.95]{USCountyHierarchy1.pdf}

\includegraphics[page=14,scale=0.95]{USCountyHierarchy1.pdf}

\includegraphics[page=15,scale=0.95]{USCountyHierarchy1.pdf}

\includegraphics[page=16,scale=0.95]{USCountyHierarchy1.pdf}

\includegraphics[page=17,scale=0.95]{USCountyHierarchy1.pdf}

\includegraphics[page=18,scale=0.95]{USCountyHierarchy1.pdf}

\includegraphics[page=19,scale=0.95]{USCountyHierarchy1.pdf}

\includegraphics[page=20,scale=0.95]{USCountyHierarchy1.pdf}

\includegraphics[page=21,scale=0.95]{USCountyHierarchy1.pdf}

\includegraphics[page=22,scale=0.95]{USCountyHierarchy1.pdf}

\includegraphics[page=23,scale=0.95]{USCountyHierarchy1.pdf}

\includegraphics[page=24,scale=0.95]{USCountyHierarchy1.pdf}

\includegraphics[page=25,scale=0.95]{USCountyHierarchy1.pdf}

\includegraphics[page=26,scale=0.95]{USCountyHierarchy1.pdf}

\includegraphics[page=27,scale=0.95]{USCountyHierarchy1.pdf}

\includegraphics[page=28,scale=0.95]{USCountyHierarchy1.pdf}

\includegraphics[page=29,scale=0.95]{USCountyHierarchy1.pdf}

\includegraphics[page=30,scale=0.95]{USCountyHierarchy1.pdf}

\includegraphics[page=31,scale=0.95]{USCountyHierarchy1.pdf}

\includegraphics[page=32,scale=0.95]{USCountyHierarchy1.pdf}

\includegraphics[page=33,scale=0.95]{USCountyHierarchy1.pdf}

\includegraphics[page=34,scale=0.95]{USCountyHierarchy1.pdf}

\includegraphics[page=35,scale=0.95]{USCountyHierarchy1.pdf}

\includegraphics[page=36,scale=0.95]{USCountyHierarchy1.pdf}

\includegraphics[page=37,scale=0.95]{USCountyHierarchy1.pdf}

\includegraphics[page=38,scale=0.95]{USCountyHierarchy1.pdf}
\section{Discussion} \label{Discussion}
For the convenience of the reader, we extract (with slight editing) 
from \cite[sec.~V.A]{SlaterDendrogram}, "Hubs and Clusters in the Evolving U. S. Internal Migration Network", a limited number of passages and tables (Tables I-II) describing interesting features--such as "cosmopolitanism" and cluster-analytic isolation indices 
\cite{slater1985cluster,ling1973probability}--of the 1995-2000 dendrograms.
\subsection{Cosmopolitanism Ranking}
The leading cosmopolitan (broad migration-base) counties found (and some of their apparently relevant features), 
in decreasing order, are:

(1) Brevard, FL (the ``Space Coast'', the Kennedy Space Center);

(2) Mohave, AZ (Lake Havasu, Grand Canyon);

(3) Clark, NV (Las Vegas);

(4) Hillsborough and Pinellas, FL, which are grouped with 
     the pair, Pasco and 
     Hernando, FL. (This 
     quartet--having an isolation index of 11.9717--is completely 
     coterminous with the governmentally designated 
     Tampa-St. Petersburg-Clearwater Metropolitan 
     Statistical Area [MSA]. Additionally, 
     Pasco and Hernando have the greatest isolation
     index, 14.6413, of any pair in the entire analysis);

(5) The southern Gulf Coast dyad formed by
    Collier County (East Naples) and Lee County (Fort Myers, a single-county 
    MSA),  FL;
    
    (6) San Diego, CA;

(7) Dallas, TX;

(8) Maricopa, AZ (Phoenix);

(9) Cook, IL (Chicago);

(10) Orange, Seminole, and Osceola, FL 
     (these three counties, along with Lake County, form the Orlando-Kissimmee
      MSA);

(12) Sumter and Lake, FL;

(13) Monroe, FL (Key West);

(14) Cumberland, NC (giant Fort Bragg and Pope Air Force Base);

(15) Mecklenburg, NC (Charlotte);

(16) Martin, St. Lucie and Indian River, FL (the lower box 
     containing ``3'') (Indian River borders Brevard 
     County, the most cosmopolitan nationally);

(17) Palm Beach, FL together with the pair
      Miami-Dade and Broward, FL;
      (this southeastern Florida 
      triad comprises the Miami-Fort Lauderdale-Pompano Beach
      MSA, highlighted in gray in the master dendrograms);

(18) Hennepin, MN (Minneapolis);

(19) Marion, FL (bordering the (17) cluster on the north);

(20) Bell, TX (Fort Hood);

(21) Polk, FL (Lakeland);

(22) Citrus, FL (formerly part of Hernando County);

(23) Weld, CO (Greeley);

(24) Larimer, CO (Fort Collins);

(25) Kenai Peninsula, AK (Seward);

(26) Los Angeles, CA; and

(27) Pierce, WA (Fort Lewis and McChord Air Force Base).

\subsection{Table of regions and their associated isolation indices}
The page references in the Table are to the ordinal-scale dendrogram.

\begin{table} \label{Table1}
\begin{tabular}{r| r | r |  c | c | c}
Region & States & Page & no. counties & i (1995-00) & i (1965-70) \\
\hline
\hline
South Jersey & NJ &6 & 7 &28.7301  & 20.8996  \\
Glades + Hendry + Okeechobee &FL & 1 & 3 & 23.474 &  \\
``Delmar'' + Baltimore & DE,MD & 5 & 15 & 20.283 & \\
Western Ohio + Randolph, IN &OH,IN & 25  & 14 & 20.0938 & \\
Western New York & NY & 7 & 18 & 19.4948 & \\
Rhode Island + S. E. Mass. & RI,MA & 3 & 12 &  18.6991 & \\
Greater Orlando& FL & 1 & 3 & 17.6523 & \\
Northern Lower Michigan & MI &8,9 & 26 & 17.2098 & \\
French Louisiana & LA &30,31 & 27 & 16.7764 & \\
Brevard & FL & 1 & 1 & 16.3097 & 19.6942 \\
Golden Triangle (Beaumont +) &TX& 4 & 6 & 16.1803 \\
Connecticut & CT &2 & 8 & 16.1339 & 25.3175 \\
Mohave (Kingman) & AZ & 1 & 1 & 15.463 & 6.39121 \\
Clark (Las Vegas) & NV &1  &1 & 15.1784 & 6.23128 \\
Rexburg, ID + Jackson, WY MSAs & ID,WY & 5 & 4 & 15.0882&  \\
Eastern Rust Belt &NJ,OH,PA,WV&24 &82 & 15.0412 & \\
Burley MSA & ID & 2  & 2 & 14.8809 & \\
Pasco + Hernando & FL &1 &2 & 14.6413 & \\
San Diego & CA & 1 & 1 & 14.2408  & 12.5938 \\
Maysville MSA + 3 counties & KY &19 & 5 & 14.1822 & \\
Hawaii & HI & 2 & 5\footnotemark[1]  & 14.121 & 12.21 \\
Northern High Plains & MT,ND,NE,SD&36,37&55&13.8799& \\
Middle Ohio Valley & IN,KY&24,25&27&13.821 \\
Eastern Shore & VA & 3 &2 & 13.7051 & \\
\hline
\hline
\end{tabular}
\caption{Most well-defined 1995-2000 migration regions and their isolation indices}
\footnotetext[1]{A fifth county, Kalawao, was included in the 
1995-00 data, but not in 1965-70}
\end{table}
\begin{table} \label{Table2}
\begin{tabular}{r| r | r |  c | c | c}
Region & States & Page & no. counties & i (1995-00) & i (1965-70) \\
\hline
\hline
Dallas & TX & 1 & 1 & 13.5473 & 14.8557 \\
Maine + 7 NH counties & ME,NH & 8  & 22 & 13.4716 & \\
Southeastern  Arizona & AZ& 2 &3 & 13.3503 & \\
Maricopa (Phoenix) & AZ & 1 & 1 & 13.2608 & 12.5479 \\
Eastern Upstate New York & NY & 7 & 28 & 13.3052 & \\
Michigan Thumb & MI & 6 & 6 & 13.2208 & \\
Wasatch Back &UT &11 &8& 13.1616 & \\
N. Vermont + Coos, NH &NH,VT& 11&10&13.0778& \\
S. Central Tennessee & TN & 22  & 10 & 13.3092 & \\
Northeast South Carolina& SC & 15  & 8 & 13.0276 & \\
Northern New England & MA,ME,NH,VT&9,10&42 & 12.8446 & \\
Cook (Chicago) & IL & 1 & 1 & 12.7682 & 16.8933 \\
Southeastern Indiana & IN & 25 & 10 & 12.7172 & \\
Northwestern Lower Michigan& MI & 9,10& 9 & 12.6567 & \\
High Colorado Rockies & CO & 3 & 3 & 12.5892 & \\
Joplin Area & MO & 5  & 3 & 12.3071 & \\
Central Savannah River & GA & 22 & 4 & 12.2086 & \\
Southern Maryland & MD & 3  & 3 & 12.1217 & \\
Amarillo (Potter + Randall) & TX & 1  &2 & 12.0528 & 8.16948 \\
Tampa MSA & FL & 1 &4 & 11.9717 & \\
York+Adams & PA & 3& 2 & 11.9433 & 13.7789 \\
Lake + Sumter & FL & 1  & 2 & 11.8635 & \\
Rhode Island & RI &  3  & 5 & 11.8384 & 11.7668\footnotemark[1]\\
Central Appalachia & MD,NC,TN,VA,WV&27,28&77& 11.7459 & \\
\hline
\hline
\end{tabular}
\caption{Most well-defined 1995-2000 migration regions and their isolation indices (cont.)}
\footnotetext[1]{Newport County was not directly clustered with the other four 
counties of the state in 1965-70}
\end{table}

\section{Cardinal-scale Dendrogram} \label{Cardinal}
\includegraphics[page=1,scale=0.95]{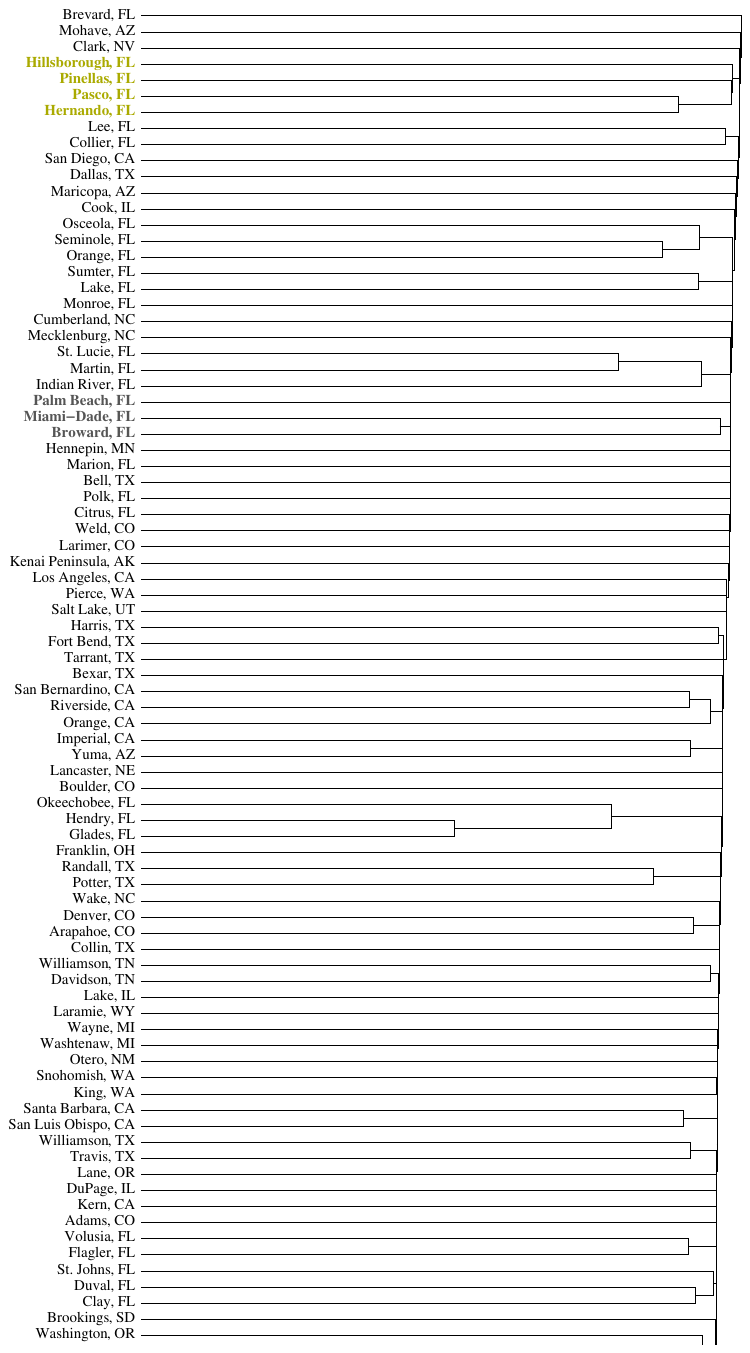}

\includegraphics[page=2,scale=0.95]{CardinalHierarchy.pdf}

\includegraphics[page=3,scale=0.95]{CardinalHierarchy.pdf}

\includegraphics[page=4,scale=0.95]{CardinalHierarchy.pdf}

\includegraphics[page=5,scale=0.95]{CardinalHierarchy.pdf}

\includegraphics[page=6,scale=0.95]{CardinalHierarchy.pdf}

\includegraphics[page=7,scale=0.95]{CardinalHierarchy.pdf}

\includegraphics[page=8,scale=0.95]{CardinalHierarchy.pdf}

\includegraphics[page=9,scale=0.95]{CardinalHierarchy.pdf}

\includegraphics[page=10,scale=0.95]{CardinalHierarchy.pdf}

\includegraphics[page=11,scale=0.95]{CardinalHierarchy.pdf}

\includegraphics[page=12,scale=0.95]{CardinalHierarchy.pdf}

\includegraphics[page=13,scale=0.95]{CardinalHierarchy.pdf}

\includegraphics[page=14,scale=0.95]{CardinalHierarchy.pdf}

\includegraphics[page=15,scale=0.95]{CardinalHierarchy.pdf}

\includegraphics[page=16,scale=0.95]{CardinalHierarchy.pdf}

\includegraphics[page=17,scale=0.95]{CardinalHierarchy.pdf}

\includegraphics[page=18,scale=0.95]{CardinalHierarchy.pdf}

\includegraphics[page=19,scale=0.95]{CardinalHierarchy.pdf}

\includegraphics[page=20,scale=0.95]{CardinalHierarchy.pdf}

\includegraphics[page=21,scale=0.95]{CardinalHierarchy.pdf}

\includegraphics[page=22,scale=0.95]{CardinalHierarchy.pdf}

\includegraphics[page=23,scale=0.95]{CardinalHierarchy.pdf}

\includegraphics[page=24,scale=0.95]{CardinalHierarchy.pdf}

\includegraphics[page=25,scale=0.95]{CardinalHierarchy.pdf}

\includegraphics[page=26,scale=0.95]{CardinalHierarchy.pdf}

\includegraphics[page=27,scale=0.95]{CardinalHierarchy.pdf}

\includegraphics[page=28,scale=0.95]{CardinalHierarchy.pdf}

\includegraphics[page=29,scale=0.95]{CardinalHierarchy.pdf}

\includegraphics[page=30,scale=0.95]{CardinalHierarchy.pdf}

\includegraphics[page=31,scale=0.95]{CardinalHierarchy.pdf}

\includegraphics[page=32,scale=0.95]{CardinalHierarchy.pdf}

\includegraphics[page=33,scale=0.95]{CardinalHierarchy.pdf}

\includegraphics[page=34,scale=0.95]{CardinalHierarchy.pdf}

\includegraphics[page=35,scale=0.95]{CardinalHierarchy.pdf}

\includegraphics[page=36,scale=0.95]{CardinalHierarchy.pdf}

\includegraphics[page=37,scale=0.95]{CardinalHierarchy.pdf}

\includegraphics[page=38,scale=0.95]{CardinalHierarchy.pdf}

\begin{acknowledgments}
This research was supported by the National Science Foundation under Grant No. NSF PHY-1748958. Dan Montello catalyzed the issuance of this report by referring me to the Jon Bruner (Forbes) website http://www.forbes.com/special-report/2011/migration.html, which presented an interactive map of American migration, based on Internal Revenue Service data.
\end{acknowledgments}

\bibliography{main}

\end{document}